\documentclass[a4paper,twoside,english,twocolumn,pra,aps,showpacs]{revtex4}
\usepackage{graphicx}
\usepackage{amssymb}
\usepackage{bm}

\makeatletter

\makeatletter
\usepackage{mathrsfs}

\newcommand{\varX}[0]{\ensuremath{\mathbf{\mathfrak X}}}

\makeatother

\makeatother

\makeatother

\makeatother

\usepackage{babel}
\makeatother
\begin{document}

\title{Quantum limits to center-of-mass measurements}

\author{Timothy Vaughan}

\email{vaughan@physics.uq.edu.au}

\affiliation{Australian Centre of Quantum-Atom Optics, University of Queensland,
St Lucia, Queensland, Australia.}

\author{Peter Drummond}

\affiliation{Australian Centre of Quantum-Atom Optics, University of Queensland,
St Lucia, Queensland, Australia.}

\author{Gerd Leuchs}

\affiliation{Institut für Optik, Information und Photonik, Max-Planck Forschungsgruppe,
Universität Erlangen-Nürnberg, Günther-Scharowsky-Straße 1/Bau 24,
91058 Erlangen, Germany}

\date{\today{}}

\begin{abstract}
We discuss the issue of measuring the mean position (center-of-mass)
of a group of bosonic or fermionic quantum particles, including particle
number fluctuations. We introduce a standard quantum limit for these
measurements at ultra-low temperatures, and discuss this limit in
the context of both photons and ultra-cold atoms. In the case of fermions,
we present evidence that the Pauli exclusion principle has a strongly
beneficial effect, giving rise to a $1/N$ scaling in the position
standard-deviation -- as opposed to a $1/\sqrt{N}$ scaling for bosons.
The difference between the actual mean-position fluctuation and this
limit is evidence for quantum wave-packet spreading in the center-of-mass.
This macroscopic quantum effect cannot be readily observed for non-interacting
particles, due to classical pulse broadening. For this reason, we
also study the evolution of photonic and matter-wave solitons, where
classical dispersion is suppressed. In the photonic case, we show
that the intrinsic quantum diffusion of the mean position can contribute
significantly to uncertainties in soliton pulse arrival times. We
also discuss ways in which the relatively long lifetimes of attractive
bosons in matter-wave solitons may be used to demonstrate quantum
interference between massive objects composed of thousands of particles. 
\end{abstract}


\maketitle

\section{Introduction}

The topic of mesoscopic quantum effects is of great current interest,
especially in photonic or in ultra-cold atomic systems where a large
decoupling from the environment is possible. An important degree of
freedom is the average or center-of-mass position of a large number
of particles. It is now possible to observe quantum diffraction of
the center-of-mass in molecules like C$_{60}$ and related fullerenes\cite{Zeil}.
Larger physical systems would provide an even a stronger test of quantum
theoretical predictions. As well as testing quantum theory in new,
mesoscopic regimes, these types of experiment have potential applications
to novel sensors. For example, ultra-precise measurements of position
may be useful in measuring gravitational interactions, or other forces
that couple to conserved quantities.

In this paper, we wish to examine the quantum limits to center-of-mass
position fluctuations. We further discuss these limits in the context
of mesoscopic systems of interacting particles, atoms or molecules.
This is an important issue in quantum-optical or ultra-cold atom environments,
where center-of-mass motion places a limit on coherence properties
of photonic or atom lasers. A localized system of particles features
a dual particle and wave nature, that is, it has conjugate observables
of momentum and position, as well as conjugate observables of number
and phase. As pointed out by Landau and Peierls\cite{Landau}, free
particle momentum is a QND observable. In fact, the momentum of a
quantum soliton can be non-destructively measured via the position
of a probe soliton after their collision\cite{Watanabe}. The quantum
position fluctuation increases with propagation distance due to an
initial momentum uncertainty. This places a fundamental limit on momentum
QND measurements.

We generalize the concept of the standard quantum limit of a mean
position measurement \cite{Lai,FiniHagelstein} to any kind of quantum
field. The standard quantum limit is defined here as the position
uncertainty of a many-body ground-state for non-interacting particles
with a given density distribution and total momentum. For massive
particles, this corresponds to the Heisenberg-limited uncertainty
of the center-of-mass of a group of non-interacting particles in an
external potential, at zero temperature. Intriguingly, we find completely
different scaling laws for different particle statistics: while bosons
have a position uncertainty (standard deviation) that scales as $1/\sqrt{N}$,
fermionic center-of-mass uncertainties scale as $1/N$ for $N$ particles.
This is caused by the particle correlations induced by Fermi statistics.
The scaling indicates that ultra-cold fermions are the preferred system
for ultra-sensitive measurements of center-of mass.

A necessary requirement for quantum-limited measurements is an extremely
low level of fluctuations and noise, which makes laser pulses and
ultra-cold atoms the most reasonable choices at present. As an example,
the localized solitons of the attractive one-dimensional Bose gas
can now be observed both in quantum-optical and in atom-optical environments.
As well as technological applications in communications, these types
of soliton show intrinsic quantum effects. That is, the effects of
quantum phase diffusion (squeezing)\cite{Carter,Lai,DrCorney} have
been observed in experiments\cite{Rosenbluh,Nature,CorneyLeuchs}.
Quantum correlations established by soliton collision (quantum non-demolition
measurement)\cite{Watanabe} have also been demonstrated in experiment\cite{Friberg}.
Even though they involve up to $10^{9}$ particles, these effects
simply do not occur in classical soliton theory. 

Phase squeezing or phase QND measurement does not lead to center-of-mass
changes, which are especially interesting for massive particles, as
this degree of freedom is coupled directly to the gravity field. Due
to problems in quantizing gravity, there have been suggestions that
gravitational effects may lead to wave-packet collapse and/or dissipation\cite{Diosi,Ghirardi,Penrose,Tod,Filippo}.
While this remains speculative, it is clearly an area of quantum mechanics
where there are no existing tests. To exclude such theories, one would
need to violate a classical inequality involving mesoscopic numbers
of massive particles, whose center-of-mass was in some type of quantum
superposition state\cite{ReidCaval}. In real experiments, weak localization\cite{Aharonov}
can take place due to interactions with the environment. It is at
least interesting to investigate how this may occur, as a first step
towards testing more fundamental types of localization. We note that
steps in this direction have been taken recently for non-interacting,
massless particles, via position observations with photons, at the
standard quantum limit or better\cite{Trepsetal,Kolobov}.

Quantum wave-packet spreading of the center-of-mass is an intrinsic
quantum prediction for untrapped particles at all particle numbers.
However, this is masked by the single-particle effects analogous to
classical pulse broadening for an optical pulse in a linear dispersive
medium. A soliton or quantum bound-state can suppress this single-particle
or classical pulse broadening, and allows the intrinsic center-of-mass
quantum spreading to be observed \cite{FiniHagelstein}, in a similar
way to the known quantum noise effects found in amplified solitons\cite{Gordon}.
We find that this is an exactly soluble problem for \textit{any} initial
quantum state. No linearization or decorrelation approximation is
needed, which allows us to examine quantum wave-packet spreading over
arbitrary distance scales. In order to distinguish classical pulse
broadening from quantum wave-packet spreading, we calculate the standard
quantum limit of this measurement for an initially coherent soliton
pulse. This requires knowledge of the time-evolution of the density
distribution for an interacting many-body system, which is nontrivial,
and requires numerical solutions.

\section{Center of mass operators}

We start by noting that while a centroid exists for any localized,
measurable quantity, it is most interesting for a conserved quantity.
For any physical system with a conserved current four-density $\widehat{j}=\left(j_{0,}\mathbf{j}\right)$,
it is possible to define a center-of-charge (or mass, energy, etc)
relative to the conserved quantities \begin{equation}
\widehat{J}_{\mu}=\int\hat{j}_{\mu}\left(\mathbf{x}\right)d^{D}\mathbf{x}\,\,.\end{equation}
 In a quantum state that is an eigenstate of the conserved charge
with $J_{0}>0$, defined on a $D$- dimensional space, the definition
of the centroid is: \begin{equation}
\hat{\mathbf{x}}_{J}=\frac{1}{J_{0}}\int\mathbf{x}\hat{j}_{0}\left(\mathbf{x}\right)d^{D}\mathbf{x}\,\,.\end{equation}

After partial integration, the conservation law, \begin{equation}
\frac{\partial\widehat{j}_{0}}{\partial t}+\nabla\cdot\hat{\mathbf{j}}=0\,\,,\end{equation}
 then implies that the centroid position translates with a velocity
given by the spatial part of the conserved current:

\begin{equation}
\frac{\partial\hat{\mathbf{x}}_{J}}{\partial t}=\frac{\hat{\mathbf{J}}}{J_{0}}\,\,.\label{eq:COMevolve}\end{equation}

While the concept of measuring the center of mass position is well-known,
there are a number of issues involved. The most obvious is that there
is a difference between the center-of-mass and the average position
of a group of non-identical particles. This difference vanishes for
non-relativistic particles all of the same mass: for example, isotopically
pure, ultra-cold atoms. However, there is a real difference in the
case of particles of different masses, and for extremely relativistic
particles and photons, where the effective mass depends on the momentum.
Here the average position differs from the center-of-mass. Another
subtlety which this paper treats in some detail, occurs when the system
is in a quantum mixture of different particle numbers: the effects
of this are treated in the remainder of this section.

\subsection{Massive fields}

\label{sub:Massive-fields}We first consider non-relativistic massive
quantum fields $\hat{\Psi}_{i}\left(\mathbf{x}\right)$, where each
component has an equal mass. The fundamental particle statistics can
be either fermionic or bosonic. As usual for quantum fields, the commutators
are: \begin{equation}
\left[\hat{\Psi_{i}}\left(t,\mathbf{x}\right),\hat{\Psi}_{j}^{\dagger}\left(t,\mathbf{x}'\right)\right]_{\pm}=\delta_{ij}\delta^{3}\left(\mathbf{x}-\mathbf{x}'\right).\end{equation}

Introducing the particle density \begin{equation}
\hat{n}\left(\mathbf{x}\right)=\sum_{i}\hat{n}_{i}\left(\mathbf{x}\right)=\sum_{i}\Psi_{i}^{\dagger}\left(\mathbf{x}\right)\hat{\Psi}_{i}\left(\mathbf{x}\right),\end{equation}
 we can define the total particle number \begin{equation}
\hat{N}=\int d^{D}\mathbf{x}\hat{n}\left(\mathbf{x}\right),\end{equation}
 which is exactly conserved in the absence of dissipative processes
like absorption or amplification. In this paper, we will consider
cases where the system has an uncertainty in $\hat{N}$, but we will
exclude states with zero particle number. In terms of operational
measurements with small particle numbers, this may require post-selection
to exclude states of this type - since of course the center-of-mass
is undefined in such cases.

In order to define center-of-mass or mean position for systems with
particle number fluctuations, several definitions are possible, depending
on measurement procedures. There is more than one measurement procedure
possible, since the quantum state need not be an eigenstate of the
number operator. Hence, different number states could have an arbitrary
relative weighting.

In the absence of external potentials, systems of interacting particles
also have an invariant quantity due to translational invariance, which
is the total momentum (in $D$ dimensions): \begin{equation}
\hat{\mathbf{P}}=\frac{\hbar}{2i}\int\left[\hat{\Psi}_{i}^{\dagger}\left(\mathbf{x}\right)\mathbf{\nabla}\hat{\Psi}_{i}\left(\mathbf{x}\right)-\nabla\hat{\Psi}_{i}^{\dagger}\left(\mathbf{x}\right)\hat{\Psi}_{i}\left(\mathbf{x}\right)\right]d^{D}\mathbf{x}\end{equation}
 Here we use the Einstein summation convention for repeated indices.
We use a capital letter to emphasize that this is an extensive quantity,
proportional to the number of particles.

\subsection{Massless fields}

Massless fields - photons - are the most commonly used particles where
measurements are made at the quantum level. In this case, we can distinguish
two different important cases. 

\begin{itemize}
\item Narrow band fields in one-dimensional wave-guides or fibers experience
dispersion, which causes them to behave as massive particles. In a
similar way, the paraxial approximation for beams allows diffractive
behavior to be treated in an effective mass approximation. These
cases can be handled in the same way as the treatment given above.
\item More fundamental issues arise in free-space, where there is a long-standing
fundamental problem of how to define photon position. In addition,
as there is no photon mass, even the idea of center-of-mass requires
care. Instead, one must either use the concept of a center-of-energy,
or of a center of photon-number. We shall focus on the latter concept
here.
\end{itemize}
To treat the free-space average position, we can use a conserved local
density obtained from the dual symmetry properties of Maxwell equations\cite{Dual}.
We introduce $\hat{\mathbf{\mathcal{E}}}_{\sigma}$, where $\sigma=\pm1$,
which are the helicity components of the complex Maxwell fields in
units where $c=\varepsilon_{0}=\mu_{0}=1$, and the corresponding
complex vector potentials $\hat{\mathbf{\mathcal{A}}}_{\sigma}$ ,
so that:\begin{eqnarray*}
\hat{\mathbf{\mathcal{E}}} & = & \hat{\mathbf{E}}+i\hat{\mathbf{B}}\\
 & = & \nabla\times\hat{\mathbf{\mathcal{A}}}\\
 & = & \hat{\mathbf{\mathcal{E}}}_{+}+\hat{\mathbf{\mathcal{E}}}_{-}\,\,.\end{eqnarray*}

It is easily checked from the complex form of Maxwell's equations
that $i\partial\hat{\mathbf{\mathcal{E}}}_{\sigma}/\partial t=\nabla\times\hat{\mathbf{\mathcal{E}}}_{\sigma}$and
$i\partial\hat{\mathbf{\mathcal{A}}}_{\sigma}/\partial t=\nabla\times\hat{\mathbf{\mathcal{A}}}_{\sigma}$.
The photon density is then: \begin{equation}
\hat{n}\left(\mathbf{x}\right)=\sum_{\sigma}\hat{n}_{\sigma}\left(\mathbf{x}\right)=\frac{\sigma}{4}:\left(\hat{\mathbf{\mathcal{E}}}_{\sigma}^{\dagger}\cdot\hat{\mathbf{\mathcal{A}}}_{\sigma}+\hat{\mathbf{\mathcal{E}}}_{\sigma}\cdot\hat{\mathbf{\mathcal{A}}}_{\sigma}^{\dagger}\right):\,\,,\label{photondensity}\end{equation}
with a corresponding conserved current of :

\begin{equation}
\widehat{\bm{J}}=\sum_{\sigma}\frac{\sigma}{4i}:\left(\hat{\mathbf{\mathcal{E}}}_{\sigma}^{\dagger}\times\hat{\mathbf{\mathcal{A}}}_{\sigma}+\hat{\mathbf{\mathcal{A}}}_{\sigma}^{\dagger}\times\hat{\mathbf{\mathcal{E}}}_{\sigma}\right)\,\,.\label{photoncurrent}\end{equation}

We note that with these definitions, the photon density is not positive
definite in small volumes and time-intervals. However, the total photon
number $\widehat{N}$ is well-defined and positive definite, so that
a position centroid is well-defined for this conserved quantity. For
the remainder of this paper we will focus on non-relativistic massive
particles for definiteness, while noting that many results that only
depend on the existence of a conserved current will hold in the general
case of a photon field.

\subsection{Intensive position }

For quantum states that \textit{are} eigenstates of number with $N>0$,
the obvious definition is: \begin{equation}
\hat{\mathbf{x}}^{(N)}=\frac{\int\mathbf{x}\hat{n}\left(\mathbf{x}\right)d^{D}\mathbf{x}}{N}\end{equation}
 This is undefined for the vacuum state, but is well-defined for all
other number eigenstates. For the non-relativistic massive particle
case, this obeys the expected commutation relation with the total
momentum operator:

\[
\left[\hat{x}_{i}^{(N)},\hat{P}_{j}\right]_{\pm}=i{\hbar\delta}_{ij}.\]

The presence of the vacuum in states composed of mixtures or superpositions
of number eigenstates introduces the requirement of deciding which
relative weights to use for different particle numbers. The important
fact to note is that the mean position of any vacuum component is
\emph{undefined} and, as such, its presence does not contribute to
one's knowledge of the mean position. Accordingly, as mentioned previously,
we choose to work in a restricted Hilbert space which does not include
the vacuum state. Practically, this is equivalent to performing heralded
particle number measurements where any null results are discarded.
From a mathematical perspective, we work with a projected density
operator $\rho'=\hat{\mathcal{P}}\rho\hat{\mathcal{P}}$, where $\hat{\mathcal{P}}=(\hat{I}-|0\rangle\langle0|)$.
In order to extract position information from linear superpositions
or mixtures of number states, we introduce the center-of-mass position
operator%
\footnote{Similar mean position operators have already been proposed although
without considering the effects of number fluctuations. See, for example,
Refs.~\cite{Yao,Lai}.%
} which is well defined in our projected Hilbert space: \begin{equation}
\hat{\mathbf{x}}=\frac{\int\mathbf{x}\hat{n}\left(\mathbf{x}\right)d^{D}\mathbf{x}}{\hat{N}}\end{equation}
 This intensive quantity also obeys the expected commutation relation
with $\hat{\mathbf{P}}$: \begin{equation}
\left[\hat{x}_{i},\hat{P}_{j}\right]_{\pm}=i\hbar\delta_{ij}\end{equation}
 Thus, provided any measurement of the vacuum is discarded, $\hat{\mathbf{x}}$
and $\hat{\mathbf{P}}$ form a pair of canonical conjugate variables
with the dimensionless uncertainty relation \begin{equation}
\langle\Delta\hat{x}_{i}^{2}\rangle\langle\Delta\hat{P}_{i}^{2}\rangle\geq\frac{\hbar^{2}}{4}\end{equation}
 for states projected into our restricted Hilbert space.

\subsection{Extensive position}

Although $\hat{\mathbf{x}}$ has the usual definition of the mean
position, it's operational denominator makes it is difficult to measure
using many standard techniques, or even to represent in a normally
ordered form. We can also introduce an extensive position operator,
which produces a result proportional to the number of particles as
well as their position. We define: \begin{equation}
\hat{\mathbf{X}}=\hat{\mathbf{x}}\hat{N},\end{equation}
 which is well-defined without projection in the Hilbert space. However,
$\hat{\mathbf{X}}$ and $\hat{\mathbf{P}}$ do not form a pair of
canonical conjugate variables, since \begin{equation}
[\hat{X}_{i},\hat{P}_{j}]_{\pm}=i\hbar\delta_{ij}\hat{N}.\end{equation}
 Through the uncertainty relation, we see that $\hat{\mathbf{X}}$
is sensitive not only to positional fluctuations but also fluctuations
in the total particle number or beam intensity: \begin{equation}
\langle\Delta\hat{X}_{i}^{2}\rangle\langle\Delta\hat{P}_{i}^{2}\rangle\geq\frac{\hbar^{2}\langle\hat{N}^{2}\rangle}{4}.\label{eq:xbaruncertainty}\end{equation}
 Furthermore, as an \emph{extensive} operator, $\hat{\mathbf{X}}$
transforms in the following way. In a new reference frame $S'$ where
$\mathbf{x}'=\mathbf{x}+\Delta\mathbf{x}$, we find that \begin{equation}
\hat{\mathbf{X}}'=\hat{\mathbf{X}}+\Delta\mathbf{x}\hat{N}\end{equation}
 rather than just being translated by a c-number.

We note that the extensive position \emph{does} have a canonical conjugate
partner, in terms of the \emph{intensive} or mean momentum operator:\begin{equation}
\hat{\mathbf{p}}=\frac{\hat{\mathbf{P}}}{\hat{N}}\end{equation}
 with a corresponding commutator of:\begin{equation}
[\hat{X}_{i},\hat{p}_{j}]_{\pm}=i\hbar\delta_{ij}\end{equation}

\subsection{Quasi-intensive position operator}

We will finally define a quasi-intensive position operator, which
corresponds to the typical direct measurement procedures in optics.
With this operator, the position is first measured with techniques
that are extensive - but the final result is normalized by the average
particle number. This has the property that in an ensemble with variable
numbers of particles, more weight is given to those measurements that
involve the most particles. Here: \begin{equation}
\hat{\varX}=\frac{\hat{\mathbf{X}}}{\langle\hat{N}\rangle}\end{equation}
 which can still be considered a useful measure of the mean position
of a particle distribution, especially in the limit of large $\langle\hat{N}\rangle$
where total number fluctuations are suppressed. Indeed, this is the
variable that corresponds most closely to some earlier proposed operational
measures of pulse position, measured using local oscillator or homodyne
techniques. It is interesting to note that the conjugate variable,
the mean momentum, is the same as for an extensive position. This
is analogous to the mean frequency of a laser pulse in a dispersive
medium, which also corresponds closely to standard operational measurements
used in laser physics.

In order to measure fluctuations about $\langle\hat{\mathbf{X}}\rangle$
or $\langle\hat{\varX}\rangle$, however, care must be taken due to
the fact that $\hat{\varX}$ transforms in the following way under
translation: \begin{equation}
\hat{\varX}'=\hat{\varX}+\Delta\mathbf{x}\frac{\hat{N}}{\langle\hat{N}\rangle}.\end{equation}
 By considering fluctuations about $\mathbf{x}'=0$ in a new reference
frame where $\mathbf{x}'=\mathbf{x}-\langle\varX\rangle$, we define
\begin{equation}
\Delta\hat{\varX}=\hat{\varX}-\langle\hat{\varX}\rangle\frac{\hat{N}}{\langle\hat{N}\rangle}\end{equation}
 and thus minimize the number fluctuation contribution. Following
from Eq.~\ref{eq:xbaruncertainty}, this obeys the uncertainty relation\[
\langle\Delta\hat{\varX}_{i}^{2}\rangle\langle\Delta\hat{\mathbf{P}}_{i}\rangle\geq\frac{\hbar^{2}\langle\hat{N}^{2}\rangle}{4\langle\hat{N}\rangle^{2}}\]
 as the total number and momentum operators commute.

In order to avoid complications arising from this issue, we will only
consider isolated systems where it is always possible to find an inertial
reference frame in which $\langle\hat{\varX}\rangle=\langle\hat{\mathbf{P}}\rangle=0$.

\section{Quantum limits on position uncertainties }

Intrinsic quantum uncertainty in the results of projective measurements
is perhaps the most striking difference between quantum mechanical
observables and classical variables. Here we seek to characterize
this uncertainty in the center-of-mass (COM) measurement by introducing
a \emph{standard quantum limit} to the variance in the distribution
of results. This is not a lower bound - there is none - but rather
a natural limit that one can expect to achieve using standard cooling
and/or stabilization methods. The question of exactly which state
to use to calculate such a characteristic COM is difficult to answer
uniformly, especially as the states accessible to bosonic systems
are different to those accessible to fermionic systems. 

For a system with known density distribution and exchange statistics,
we \emph{define} the standard quantum limit to the COM uncertainty
to be the variance remaining when a non-interacting system of the
same particles is reduced to zero temperature in an external potential
that reproduces the given density distribution. It is important to
note that this definition makes no restriction on the statistics of
the characteristic system. However, we will find that the statistics
do have a large effect on the way that the quantum limit scales for
a given density.

In the following analysis, we consider the variance in the intensive
mean position operator, in the form:\begin{equation}
|\Delta\hat{\mathbf{x}}|^{2}=\frac{1}{\hat{N}^{2}}\int d^{D}\mathbf{x}d^{D}\mathbf{y}\left\{ \Delta\mathbf{x}\cdot\Delta\mathbf{y}\hat{n}(\mathbf{x})\hat{n}(\mathbf{y})\right\} \,\,.\end{equation}
Re-arranging this expression by using normal-ordering and commutators
gives the result that:

\begin{eqnarray}
\left|\Delta\hat{\mathbf{x}}\right|^{2} & = & \frac{1}{\hat{N}^{2}}\left(\hat{N}\hat{\sigma}^{2}+:|\Delta\hat{\mathbf{X}}|^{2}:\right)\label{eq:COMvar}\end{eqnarray}
 where we define the wavepacket variance operator \begin{equation}
\hat{\sigma}^{2}=\frac{1}{\hat{N}}\int d^{D}\mathbf{x}\left\{ \left|\Delta\mathbf{x}\right|^{2}\hat{n}(\mathbf{x})\right\} \end{equation}
 and the normally ordered COM variance \begin{eqnarray}
:|\Delta\hat{\mathbf{X}}|: & = & \int d^{D}\mathbf{x}d^{D}\mathbf{y}\Delta\mathbf{x}\cdot\Delta\mathbf{y}\nonumber \\
 &  & \times\sum_{ij}\hat{\Psi}_{i}^{\dagger}(\mathbf{y})\hat{\Psi}_{j}^{\dagger}(\mathbf{x})\hat{\Psi}_{j}(\mathbf{x})\hat{\Psi}_{i}(\mathbf{y}).\end{eqnarray}
Note that the exact operator ordering in the last expression is important
if the decomposition is to be correct for both Fermi and Bose field
operators.

\subsection{Bosonic fields}

The simplest configuration for a system of degenerate bosons has all
particles described by a single normalized mode function $\chi(\mathbf{x})$.
Thus, the characteristic states considered here are formed using functions
of the creation operator $\widehat{a}_{\chi}^{\dagger}=\int d^{D}\mathbf{x}\left\{ \chi\left(\mathbf{x}\right)\hat{\Psi}^{\dagger}\left(\mathbf{x}\right)\right\} $.

\subsubsection{Number states}

Zero temperature bosonic number states can expressed in the following
way:

\begin{equation}
\left|N\right\rangle =\frac{1}{\sqrt{N}}(\hat{a}^{\dagger})^{N}\left|0\right\rangle .\end{equation}
 Evaluating the COM variance of this state using the above decomposition
is straightforward as the expectation value of the normally ordered
term $:|\Delta\hat{\mathbf{x}}|^{2}:$ vanishes, leaving\begin{equation}
\langle|\Delta\hat{\mathbf{x}}|^{2}\rangle=\frac{\sigma^{2}}{N}\end{equation}
 where $\sigma^{2}=\langle\hat{\sigma}^{2}\rangle$.

\subsubsection{Coherent states}

For bosons, the first term also vanishes for any coherent state of
the field operators, given our co-moving reference frame. Here we
define a coherent state using a projection, as explained earlier,
which projects out the vacuum state in order to correspond to a heralded
or post-selected measurement. Such states can be formally defined
as:

\begin{equation}
\left|\alpha\right\rangle _{P}=\left[e^{\left|\alpha\right|^{2}}-1\right]^{-1/2}\sum_{n=1}^{\infty}\frac{\left[\alpha\widehat{a}_{\chi}^{\dagger}\right]^{n}}{n!}\left|0\right\rangle .\label{eq:projectedcoherent}\end{equation}

This state can be viewed as a suitable reference state for quantum
noise, in which the uncertainty in a position measurement is governed
entirely by the spread in the particle distribution, $\left|\chi\left(\mathbf{x}\right)\right|^{2}$.
After normalizing by the particle number, we find \begin{equation}
\langle{|\Delta\hat{\mathbf{x}}|}^{2}\rangle=\sigma^{2}\left\langle \frac{1}{\hat{N}}\right\rangle .\end{equation}
 This is different from the following result, obtained by considering
the variance in the quasi mean position operator $\hat{\varX}$ which
for the coherent state above is given by\begin{equation}
\langle|\Delta\hat{\varX}|^{2}\rangle=\frac{\sigma^{2}}{\langle\hat{N}\rangle}.\end{equation}

\begin{figure}[h!]
\includegraphics[width=7cm]{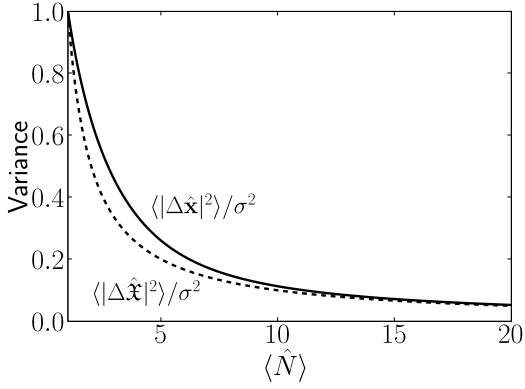}

\caption{Deviation of the variance (relative to the wavepacket variance, $\sigma^{2}$)
of the quasi intensive COM coordinate (dashed) from the true intensive
COM coordinate (solid) for a heralded coherent state.}

\label{fig:SQL} 
\end{figure}

From the graphical comparison shown as Fig.~\ref{fig:SQL} it is
clear that the standard quantum noise limit for the extensive (quasi
mean) position measurement is lower than that for the intensive position
measurement. This can be understood as a consequence of the fact that
an extensive position measurement for a coherent state effectively
weights all \emph{individual} particle position measurements equally.
This is advantageous for a coherent state - as in a laser pulse -
where all particles arrive independently, and carry the same information.
However, not all pulses have the same number of independent particles.
An intensive measurement weights all \emph{collective} particle position
measurements equally, regardless of particle number. This gives less
weight per particle measurement when the particle number is large,
and hence is less than optimal.

\subsection{Fermionic fields}

Identifying a characteristic COM uncertainty for a cold Fermi gas
is a natural extension to the above discussion. We therefore consider
a spin-polarized gas of $N$ identical fermions at zero temperature.
In other words, we consider a system in which all modes below the
characteristic Fermi energy contain a single fermion each, whilst
those above this energy are vacant. Using the Hartree-Fock assumption
of wavefunction factorization, this state can be written \begin{equation}
|\Psi^{(N)}\rangle=\left(\prod_{j=0}^{N-1}\hat{a}_{j}^{\dagger}\right)\left|0\right\rangle \end{equation}
 where the fermion creation operator for the j-th free-particle energy
level is defined as \begin{equation}
\hat{a}_{j}^{\dagger}=\int d^{D}\mathbf{x}\chi_{j}(\mathbf{x})\hat{\Psi}^{\dagger}(\mathbf{x})\end{equation}
 and $\chi_{j}(\mathbf{x})$ is an orthonormal set of mode functions
(energy eigenstate basis).

We now calculate the expectation value of the mean position variance
operator given by Eq.~\ref{eq:COMvar}. For the first term in this
equation, we employ the commutation relation $\left[\hat{\Psi}(\mathbf{x}),\hat{a}_{j}^{\dagger}\right]{}_{+}=\chi_{j}(\mathbf{x})$
to show that \begin{equation}
\hat{\Psi}(\mathbf{x})\left|\Psi^{(N)}\right\rangle =\sum_{j=0}^{N-1}(-1)^{j}\chi_{j}(\mathbf{x})\left|\Psi_{-j}^{(N)}\right\rangle \end{equation}
 where $\left|\Psi_{-j}^{(N)}\right\rangle =\left(\prod_{k=0}^{N-1}\left(\hat{a}_{k}^{\dagger}\right)^{(1-\delta_{jk})}\right)\left|0\right\rangle $.
Taking the inner product of this result with it's conjugate leads
to \begin{equation}
\sigma^{2}=\frac{1}{N}\sum_{j=0}^{N-1}\int d^{D}\mathbf{x}\left\{ \left|\Delta\mathbf{x}\right|^{2}\left|\chi_{j}(\mathbf{x})\right|\right\} .\end{equation}

To evaluate the second term in the expectation value of Eq.~\ref{eq:COMvar}
we consider the action of two Fermi field operators on the system
state vector: \begin{eqnarray}
\hat{\Psi}(\mathbf{y})\hat{\Psi}(\mathbf{x})\left|\Psi^{(N)}\right\rangle  & = & \sum_{jk=1}^{N}(-1)^{j+k-\theta(k-j)}(1-\delta_{jk})\nonumber \\
 &  & \times\chi_{j}(\mathbf{x})\chi_{k}(\mathbf{y})\left|\Psi^{(N)}\right\rangle \end{eqnarray}
 Again taking the norm of the result, we find: \begin{eqnarray}
\langle:|\mathbf{X}|^{2}:\rangle & = & \sum_{jk=1}^{N}\int d^{D}\mathbf{x}d^{D}\mathbf{y}\{\mathbf{x}\cdot\mathbf{y}[|\chi_{j}(\mathbf{x})|^{2}|\chi_{k}(\mathbf{y})|^{2}\nonumber \\
 &  & -\chi_{j}^{*}(\mathbf{x})\chi_{k}(\mathbf{x})\chi_{k}^{*}(\mathbf{y})\chi_{j}(\mathbf{y})]\}\label{eqn:novarFermi}\end{eqnarray}

Up until now, the single particle basis $\chi_{j}(\mathbf{x})$ has
been arbitrary. For definiteness, we will assume that the functions
$\chi_{j}(\mathbf{x})=\langle\mathbf{x}|j\rangle$ are energy eigenstates
of a simple harmonic oscillator of mass $m$ and frequency $\omega$.
That is, \begin{eqnarray*}
\sigma^{2} & = & \frac{1}{N}\sum_{j=0}^{N-1}\int d^{D}\mathbf{x}\{|\mathbf{x}|^{2}\langle j|\mathbf{x}\rangle\!\langle\mathbf{x}|j\rangle\}\\
 & = & \frac{1}{N}\sum_{j=0}^{N-1}\langle j|\hat{\mathbf{q}}^{2}|j\rangle,\end{eqnarray*}
where we have also defined the first quantized position operator $\hat{\mathbf{q}}=(\hat{q}_{x},\hat{q}_{y},\hat{q}_{z})$
such that $\hat{q}_{\mu}|\mathbf{x}\rangle=x_{\mu}|x\rangle$. Recalling
that this operator can also be expressed in terms of raising ($\hat{\alpha}_{\mu}^{\dagger}$)
and lowering ($\hat{\alpha}_{\mu}$) operators, we find

\begin{eqnarray*}
\sigma^{2} & = & \frac{1}{N}\left(\frac{\hbar}{2m\omega}\right)\sum_{j=0}^{N-1}\sum_{\mu=1}^{D}(\langle j|\hat{\alpha}_{\mu}\hat{\alpha}_{\mu}^{\dagger}|j\rangle+\langle j|\hat{\alpha}_{\mu}^{\dagger}\hat{\alpha}_{\mu}|j\rangle\end{eqnarray*}
and hence the variance due to the wavepacket extent is in this case\begin{equation}
\sigma^{2}=\frac{D\hbar}{2m\omega}N.\end{equation}

The normally ordered component of the COM variance can be evaluated
using a similar approach. The first line in Eq.~\ref{eqn:novarFermi}
vanishes due to the odd parity of the integrand. The remaining term
gives\begin{eqnarray*}
\langle:|\hat{\mathbf{X}}|^{2}:\rangle & = & -\sum_{jk=0}^{N-1}\sum_{\mu=1}^{D}\langle j|\hat{q}_{\mu}|k\rangle\!\langle k|\hat{q}_{\mu}|j\rangle\\
 & = & -\frac{D\hbar}{2m\omega}\sum_{jk=0}^{N-1}\left((j+1)\delta_{j+1,k}+j\delta_{j-1,k}\right)\end{eqnarray*}
 which after combining the two terms in brackets leaves\begin{eqnarray}
\langle:|\hat{\mathbf{X}}|^{2}:\rangle & = & -\frac{D\hbar}{2m\omega}N(N-1).\end{eqnarray}
Taking into account spin degeneracy of an $S$-component Fermi gas,
which gives rise to $S$ independent measurements each with $N/S$
particles, the COM variance for an $N$-particle Fermi gas at zero
temperature held in an harmonic trap is\begin{eqnarray}
\langle|\Delta\hat{\mathbf{x}}|^{2}\rangle & = & \frac{S\sigma^{2}}{N^{2}}.\end{eqnarray}

The contribution from the normally ordered term vanishes for $N=1$
and is less than zero for all $N>1$. As this term always vanishes
for a bosonic number state, while the wavepacket variance is identical
for a bosonic gas constrained to the same density profile as the fermionic
gas we have just considered (i.e. $\langle\hat{n}(\mathbf{x})\rangle=\sum_{j=1}^{N}|\chi_{j}(\mathbf{x})|^{2}$),
it is clear that the standard quantum limit for COM measurements is
intrinsically lower for fermions than for bosons. In the case of a harmonic
trap, the variance is $N$ times smaller for fermions than for bosons.

\section{Quantum wave packet spreading}

In pure state evolution, any increase in the uncertainty \emph{above}
the standard quantum limit is regarded as originating in quantum wave-packet
spreading of the center-of-mass. We thus define \begin{equation}
\sigma_{QM}=\sqrt{\langle\Delta\varX^{2}\rangle-\sigma_{SQL}^{2}}\end{equation}
 to measure this phenomenon. We note that states for which $\sigma_{QM}\gg0$
are the ``strange quantum states'' which are mentioned in the photonic
case by some previous workers\cite{FiniHagelstein}, and also play
a role in the theory of quantum non-demolition position measurements
and gravity-wave detection. This must be carefully distinguished from
uncertainties in mixed states, which are not caused by quantum superpositions.

A similar property to quantum wave-packet spreading is already known,
and observed experimentally. In the Gordon-Haus effect\cite{Gordon}
in amplifying optical fibers, the soliton time-of-arrival is perturbed
by the effect of laser amplification. Soliton timing jitter of this
type is an extrinsic property of the amplifying system. Instead, we
will treat the intrinsic quantum effects which are present even in
the absence of the spontaneous emission noise of a laser amplifier.

To solve for the time evolution of the COM uncertainty of arbitrary
states, we first note that Eq.~\ref{eq:COMevolve} can be directly
integrated to give the following result, for any (effectively non-relativistic)
system prepared in an eigenstate of the total particle number operator:\[
\hat{\mathbf{x}}_{N}(t)=\hat{\mathbf{x}}_{N}(0)+t\frac{\hat{\mathbf{P}}}{mN}.\]
where $m$ can be either the mass of an atom or the effective mass
of a polariton, as discussed in section \ref{sub:Massive-fields}
above. For extreme relativistic or massless photon propagation, the
lack of a longitudinal dispersion mechanism implies that wave-packet
spreading is analogous to diffraction, and occurs transversely to the
propagation direction. This must be treated using the conserved number
density current operator of Eq.~(\ref{photoncurrent}).

This result can be generalized to other states, provided that the
total number is conserved during the evolution. Under this condition,
the Heisenberg picture evolution of the intensive and quasi-intensive
mean position operators are respectively \begin{equation}
\hat{\mathbf{x}}(t)=\hat{\mathbf{x}}(0)+t\frac{\hat{\mathbf{P}}}{m\hat{N}},\,\,\textrm{and}\end{equation}
\begin{equation}
\hat{\varX}(t)=\hat{\varX}(0)+t\frac{\hat{\mathbf{P}}}{m\langle\hat{N}\rangle}.\end{equation}
In the pulse frame, taking the squares gives the evolution of the
(true intensive and quasi intensive) mean position:\begin{eqnarray}
\langle|\Delta\hat{\mathbf{x}}(t)|^{2}\rangle & = & \langle|\hat{\mathbf{x}}(0)|^{2}\rangle+\frac{t}{m}\langle(\hat{\mathbf{x}}(0)\cdot\hat{\mathbf{P}}+\hat{\mathbf{P}}\cdot\hat{\mathbf{x}}(0))/\hat{N}\rangle\nonumber \\
 &  & +\frac{t^{2}}{m^{2}}\langle|\hat{\mathbf{P}}|^{2}/\hat{N}^{2}\rangle,\,\,\textrm{and}\label{eq:VarEvolveTrue}\end{eqnarray}
 \begin{eqnarray}
\langle|\Delta\hat{\varX}(t)|^{2}\rangle & = & \langle|\hat{\varX}(0)|^{2}\rangle+\frac{t}{m}\langle\hat{\varX}(0)\cdot\hat{\mathbf{P}}+\hat{\mathbf{P}}\cdot\hat{\varX}(0)\rangle/\langle\hat{N}\rangle\nonumber \\
 &  & +\frac{t^{2}}{m^{2}}\langle|\hat{\mathbf{P}}|^{2}\rangle/\langle\hat{N}\rangle^{2}.\label{eq:VarEvolveQuasi}\end{eqnarray}

\subsection{Bosonic coherent state evolution}

When the system is initially prepared in a coherent state (projected
to remove the vacuum state, as in Eq.~\ref{eq:projectedcoherent})
the mean position uncertainties evolve as \begin{equation}
\langle|\Delta\hat{\varX}(t)|^{2}\rangle_{P}=\frac{1}{\langle\hat{N}\rangle_{P}}\left[\alpha+\beta\frac{\hbar t}{m}+\gamma\left(\frac{\hbar t}{m}\right)^{2}\right]\label{eq. 14}\end{equation}
 where $\alpha,\beta,\gamma$ are functions of the initial wave-packet
shape only. These are given by: \begin{eqnarray*}
\alpha & = & \int|\mathbf{x}|^{2}|\chi(\mathbf{x})|^{2}d^{D}\mathbf{x}\\
\beta & = & i\int\mathbf{x}\cdot\left[(\nabla\chi^{*}(\mathbf{x}))\chi(\mathbf{x})-\chi^{*}(\mathbf{x})(\nabla\chi(\mathbf{x}))\right]d^{D}\mathbf{x}\\
\gamma & = & \int|\nabla\chi(x)|^{2}d^{D}\mathbf{x}\end{eqnarray*}
 where $\chi(\mathbf{x})$ is again the normalized spatial mode in
the co-moving frame.

This result agrees with previous approximate linearized results obtained
for initial coherent solitons\cite{Lai}, but is much more general.
It is valid for \textit{any} initial coherent state, at large photon
number, independent of the nonlinear interaction, due to the fact
that the coefficients depend only on the initial coherent state. If
we recall that the Hamiltonian corresponds to coupled propagation
of massive bosons, then the reason for this exact independence of
the coupling is transparent. The definition of $\hat{\varX}$ corresponds
to a center-of-mass measurement, which never depends on the two-body
coupling. However, the standard quantum limit \textit{does} depend
on the nonlinearity. This is because the coupling changes the pulse-shape
$\langle\hat{\Psi}^{\dagger}(\mathbf{x})\hat{\Psi}(\mathbf{x})\rangle$,
as an initially coherent pulse with uncorrelated bosons evolves into
a correlated state. Thus, for example, if a 1D coherent source produces
a sech input pulse, so that $\chi(x)=\mbox{sech}(x/2)/2$, then $\alpha=\pi^{2}/3\,{\,\textrm{m}}^{2};\beta=0;\,\,\gamma=1/12\,\,\textrm{m}^{-2}$.
This is not a minimum uncertainty state, as: \begin{equation}
\langle\Delta{\hat{\varX}(0)}^{2}\rangle_{P}\langle\Delta\hat{P}^{2}\rangle_{c}=\pi^{2}/36=0.27415..>0.25.\end{equation}
 Coherent Gaussian input pulses, on the other hand, \textit{are} a
minimum uncertainty state in momentum and position, as one might expect.
However, these have a large continuum radiation when used to form
solitons. This leads to a paradox: the sech-type coherent soliton
has a larger uncertainty product than a coherent Gaussian pulse, yet
experimentally appears more localized!

More complex input states than coherent states can be considered,
and they also can be treated exactly. For example, Haus and Lai\cite{Lai}
have considered an input state of spatially correlated photons, consisting
of a superposition of eigenstates of the \textit{interacting} Hamiltonian.
In this case, the mean pulse-shape is a sech pulse. However, due to
boson-boson correlations, the resulting state is a minimum uncertainty
state in position and momentum\cite{Kartner}. It is unlikely that
any existing laser source can produce the required correlations. Nevertheless,
this example shows that, when dealing with correlated states, it is
possible to reach the minimum uncertainty limit with a non-Gaussian
pulse-shape.

\section{Soliton propagation}

We next consider weakly interacting particle distributions confined
to a single transverse mode of a waveguide, which can therefore be
treated by an effective one dimensional field theory. Examples of
such systems include photonic wave packets in single-mode optical
fibers and atomic BEC clouds moving in 1D wave guides. Such physical
systems have the property that they can form solitons, in which the
classical wave-packet spreading is minimized, thus forming excellent
candidates for observation of these center-of-mass uncertainties.

\subsection{Effective field theory}

Our starting point is the standard Hamiltonian operator describing
a 1D ensemble of spinless bosons which interact through a simple delta-function
potential. In terms of a dimensionless particle density amplitude,
$\hat{\phi}$, the Hamiltonian can be written: \begin{equation}
\hat{H}=\frac{1}{2}\int d\zeta\left[\hat{\phi}_{\zeta}^{\dagger}(-\partial_{\zeta}^{2})\hat{\phi}_{\zeta}\pm\hat{\phi}_{\zeta}^{\dagger2}\hat{\phi}_{\zeta}^{2}\right]\label{eq:H}\end{equation}
 where the field operators have the commutation relation \begin{equation}
[\hat{\phi}_{\zeta}(\tau),\hat{\phi}_{\zeta'}^{\dagger}(\tau)]=\frac{1}{\bar{n}}\delta(\zeta-\zeta').\end{equation}
 Heisenberg's equations of motion for the field operators are then
\begin{equation}
i\partial_{\tau}\hat{\phi}_{\zeta}=\bar{n}[\hat{\phi}_{\zeta},\hat{H}].\end{equation}
 These equations simultaneously describe at least two physically distinct
many-body systems, the details of which are implicit in the characteristic
time $t_{0}=t/\tau$ and length $x_{0}=x/\zeta$ which link the physical
and dimensionless coordinate systems. Together with the mean particle
number $\bar{n}$ and the sign of the nonlinearity, these scaling
factors completely specify a particular system.

Loss (or gain) mechanisms can be formally incorporated into this model
by including terms in the Hamiltonian which couple the system to external
degrees of freedom. After tracing over these reservoir states, this
procedure leads to a master equation for the reduced density operator
in generalized Lindblad form, which includes c-number damping coefficients.
Although these are frequently added in an \emph{ad hoc} fashion, it
is important to note that in doing so one implicitly assumes that
the reservoir is at very low temperature and has no thermal contribution
to the system modes.\cite{Carmichael}

\subsection{Numerical results}

\begin{figure}
\includegraphics[width=8cm]{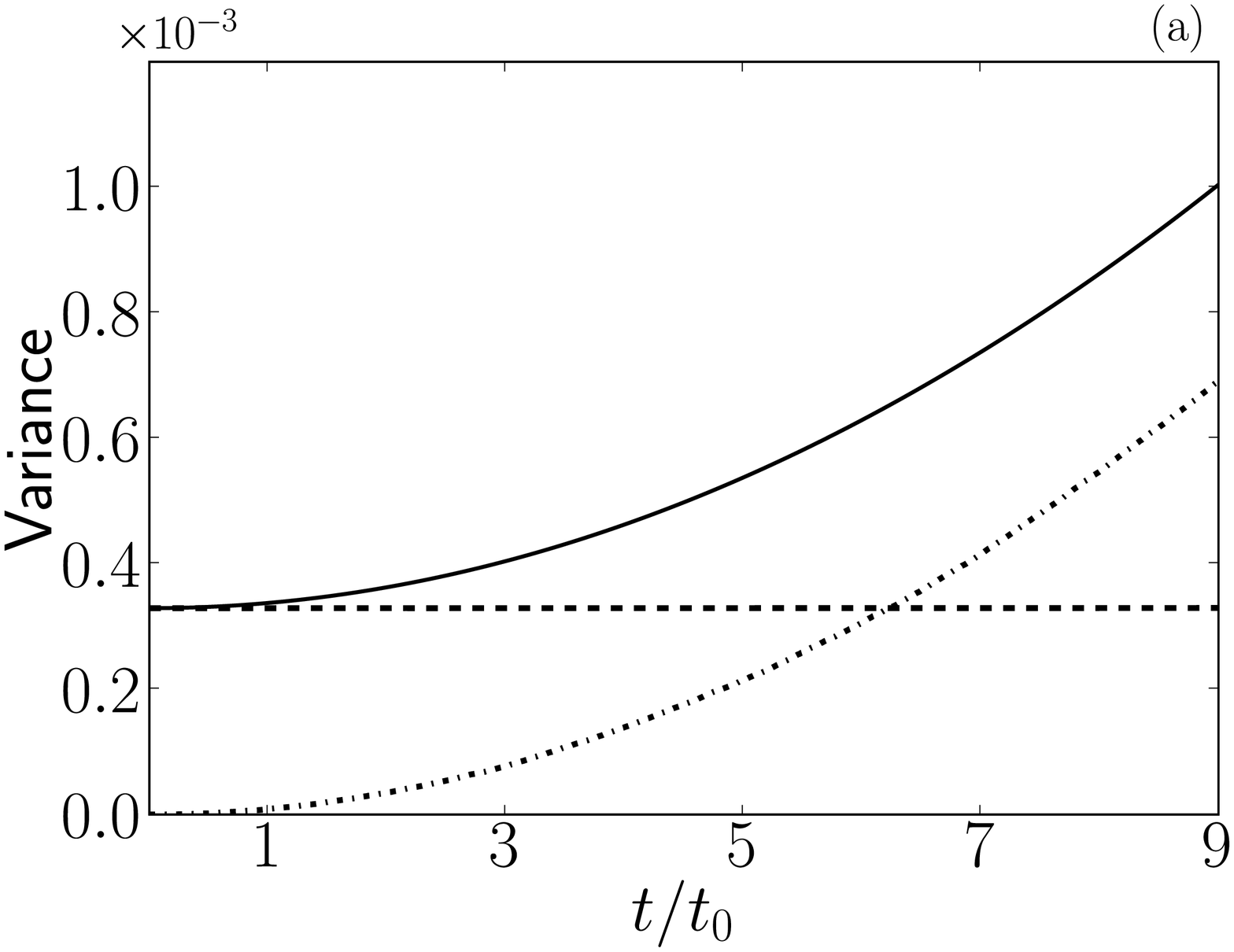}\vspace*{0.5cm}

\includegraphics[width=8cm]{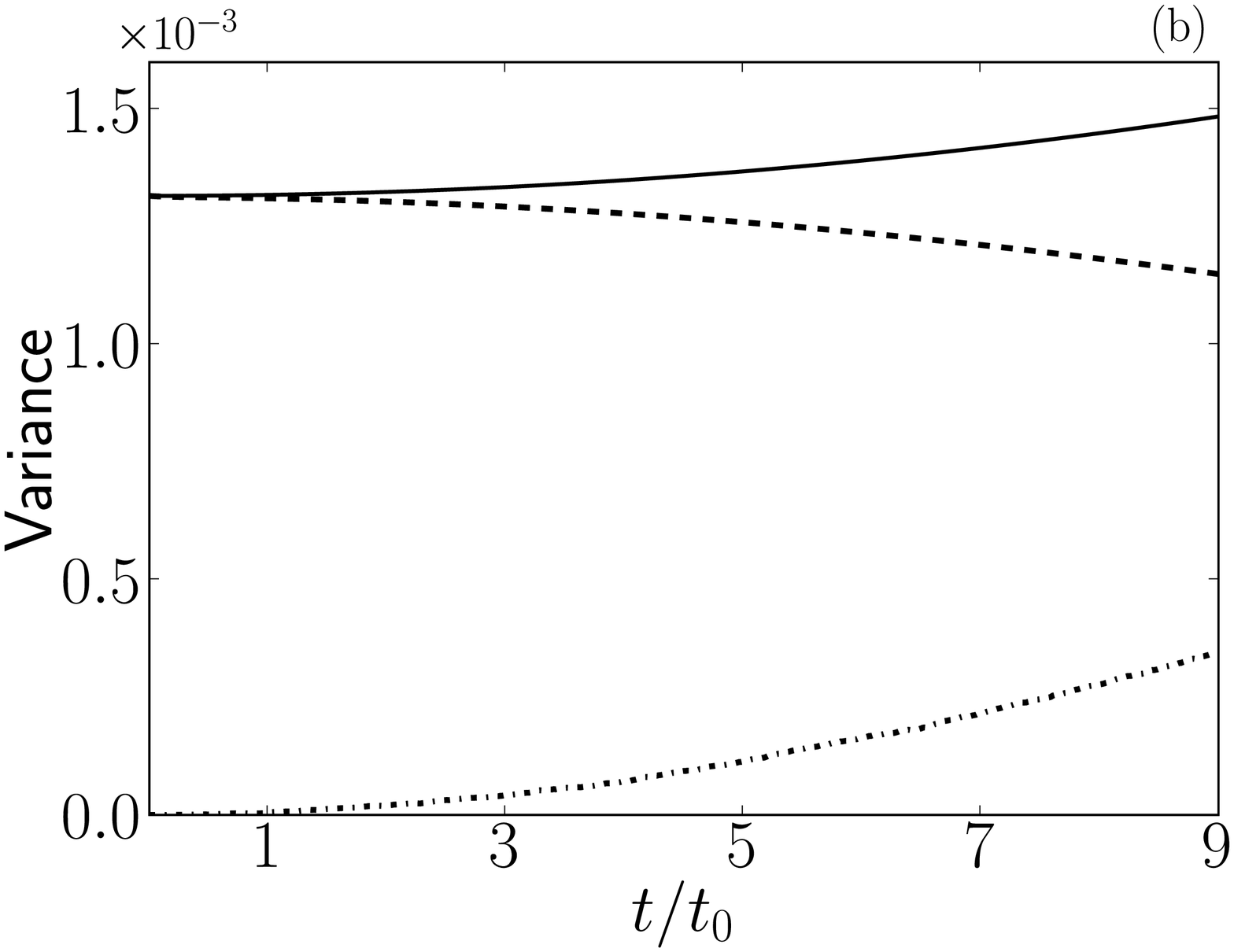}\vspace*{0.5cm}

\includegraphics[width=8cm]{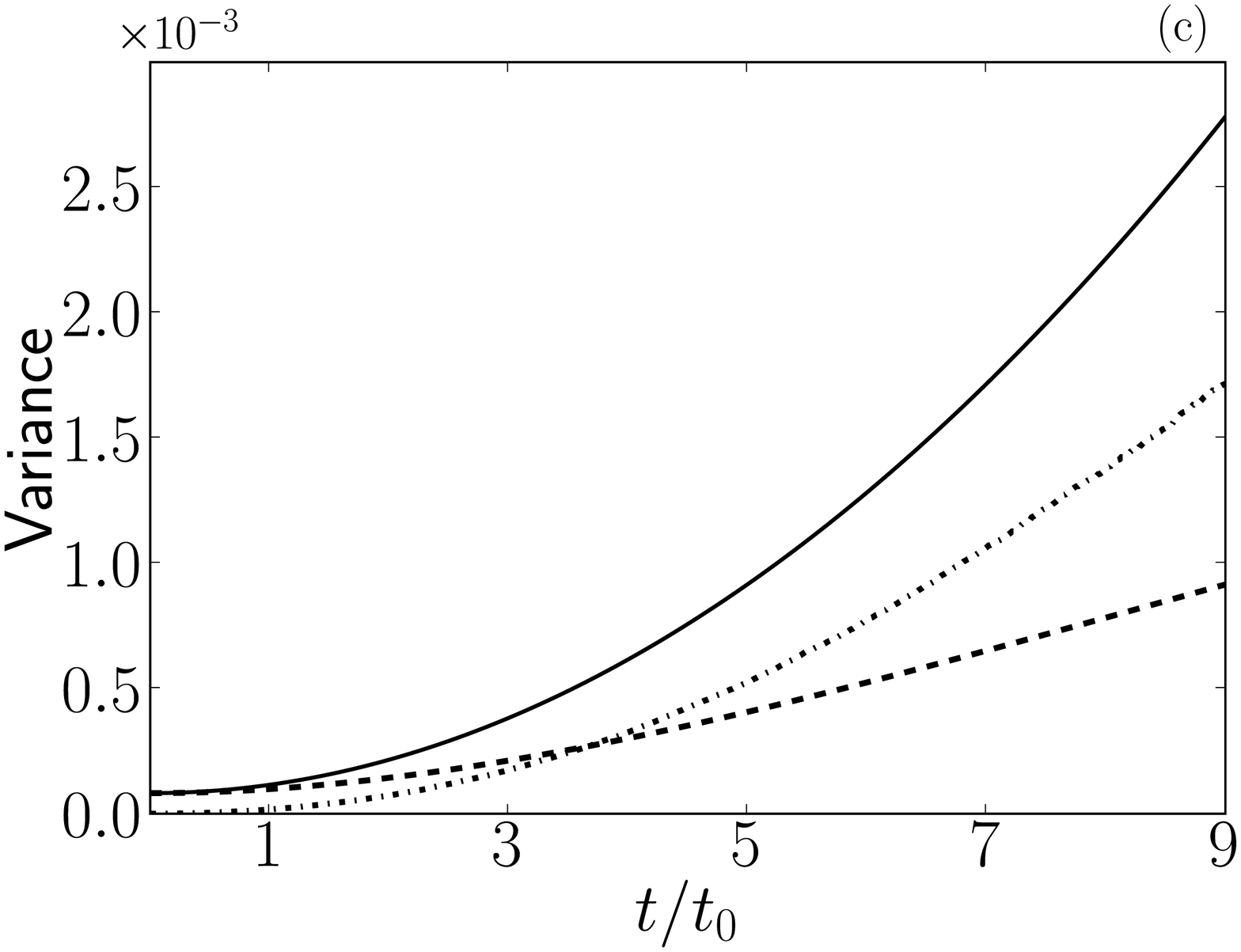}

\caption{Mean position variances for 1D coherent pulses of $\bar{n}=10^{4}$
particles propagating in a dispersive medium. The three figures correspond
to three different wavepacket shapes: (a) the soliton solution to
the 1D nonlinear Schr\"odinger equation $\langle\hat{\phi}(\zeta)\rangle=\frac{1}{2}\textrm{sech}(\frac{1}{2}\zeta)$,
(b) a wider pulse of the form $\langle\hat{\phi}(\zeta)\rangle\propto\textrm{sech}(\frac{1}{4}\zeta)$
and (c) a narrower pulse $\langle\hat{\phi}(\zeta)\rangle\propto\textrm{sech}(\zeta)$.
The solid lines represents the analytically determined total variance
$\langle\Delta\hat{\varX}^{2}\rangle/x_{0}^{2}$, while the dashed
and dash-dotted lines represent the numerical results for the classical
($\sigma_{SQL}^{2}/x_{0}^{2}$) and quantum ($\sigma_{QM}^{2}/x_{0}^{2}$)
contributions respectively.}

\label{fig:variance} 
\end{figure}

Typical results for the mean position variances during the propagation
of a $10^{4}$ particle coherent state soliton with initial shape
$\langle\hat{\phi}_{\zeta}\rangle=\frac{1}{2}\mbox{sech}(\frac{1}{2}\zeta)$
are shown in Fig.~\ref{fig:variance}(a). Here the dashed line represents
the analytic solution for the total variance $\langle\Delta\hat{\varX}^{2}\rangle/x_{0}^{2}$,
while the solid lines are numerical results generated using a stochastic
computer simulation equivalent to the quantum nonlinear Schrödinger
equation\cite{Carter}. These results include the total variance (in
complete agreement with the analytical result), the standard quantum
limit $\sigma_{SQL}^{2}$ and the measure of quantum wave packet spreading
$\sigma_{QM}^{2}$. For comparison, Figs.~\ref{fig:variance}(b)
and \ref{fig:variance}(c) display the evolution of these variances
for wider and narrower pulses than the classical soliton solution.

The standard quantum limit for the soliton is constant, which indicates
each soliton envelope remains nearly invariant; as one might expect
from a classical analysis. This is a remarkable macroscopic quantum
effect. An optical coherent state soliton consists of linear superpositions
of continuum photons and different number-momentum eigenstates. This
is clearly different to a classical soliton with frequency jitter,
yet it behaves very similarly as far as the center position is concerned.
On the other hand, the standard quantum limit for a pulse in a \textit{linear}
medium increases in the same way as the soliton quantum wave-packet
spreading, due to the dispersive spreading in the average intensity.
In the linear case, an initial coherent state remains coherent, so
the increased quantum positional uncertainty can be attributed to
the shot-noise error intrinsic to the detection of a pulse whose envelope
is not sharply localized. 

This indicates that the intrinsic quantum wave-packet spreading can
be observed only for the soliton, where it is easily distinguishable
from coherent shot-noise effects. However, this does not mean that
a quantum soliton is unstable. Each soliton preserves its soliton
pulse shape. In this sense, the soliton quantum diffusion is different
from the classical pulse broadening due to linear dispersion. In fact,
in the case of the correlated initial state, the second-order and
the fourth-order correlation functions are invariant\cite{Kousnetsov}.

\section{Practical examples}

In this section we consider some practical examples and numerical
estimates with currently available experimental technologies.

\subsection{Photonic systems}

In the optical case, the soliton propagation model describes a distribution
of interacting photons (or strictly, polaritons) propagating in a
single-mode fiber with a third-order Kerr nonlinearity, $\chi^{(3)}$.
Here the sign of the interaction term is positive for normal dispersion
and negative for anomalous dispersion. For $\chi^{(3)}$ soliton propagation,
we require the carrier frequency to be inside the anomalous dispersion
regime.

Typical orders of magnitude of soliton parameters are $\bar{n}=10^{9}$,
$t_{0}=10^{-12}$s, $x_{0}=1$m. With these parameters, we obtain
a standard quantum limit of $10^{-8}$m for a single coherent pulse.
In $1$km of propagation, the wave packet would therefore spread to
$10^{-5}$m, much greater than the standard quantum limit. Although
still less than the actual soliton extent of $10^{-4}$m, this is
a remarkably large quantum wave-packet spreading in a composite object
of $10^{9}$ particles. An experiment on quantum-mechanical wave-packet
spreading in a composite object of this type (such as that described
in \cite{FiniHagelstein}) would test quantum mechanics in a region
of much larger particle number than previously achieved.

Unfortunately, however, even fiber attenuations lower than the usual
$0.1$dB/km translate to losses of over $10^{7}$ photons from such
a pulse over this distance. Since each particle lost can be used to
obtain information about the mean position of a pulse, these losses
would destroy the coherence properties of a wave packet. Even so,
the remaining statistical uncertainty is a practical concern -- it
is, for example, likely to have a severe effect on the error-rate
in pulse-position logic\cite{Islam}. The effect may be enhanced by
using dispersion-engineered fibers\cite{FiniHagelstein}.

\subsection{Degenerate Bose gasses}

The same effective field theory can also describe a dilute, single-species
gas of massive bosonic particles, interacting through low-energy S-wave
($l=0$) scattering events and held in a potential which confines
the particles to a single transverse mode but allows longitudinal
propagation. In this case, one finds that: \begin{eqnarray}
x_{0} & = & \frac{\hbar^{2}}{m\bar{n}g_{1D}}\nonumber \\
t_{0} & = & \frac{\hbar^{3}}{m\bar{n}^{2}g_{1D}^{2}}\end{eqnarray}
 where $m$ is the single-particle mass, and $g_{1D}=2\hbar a_{s}\omega_{\perp}$
is the effective 1D interaction strength\cite{Abdullaev}. For temperatures
well below the BEC transition temperature, such gasses are usually
considered to be the matter-wave analog of coherent laser output,
and are thus sometimes referred to as atom-lasers.

Although there are various paths one can take to realizing matter-wave
solitons, we consider here only bright solitons in the absence of
any periodic potential -- i.e., solitons for which the required nonlinearity
is provided by an attractive interaction between atoms. Such solitons
have been observed\cite{Hulet,Salomon} in BECs of $^{7}$Li atoms,
where the presence of a Feshbach resonance is used to tune the effective
scattering length from repulsive to attractive. These experiments
present the very real possibility of directly observing the quantum
dispersion of a mesoscopic object composed of \emph{massive} particles,
thanks to the extremely slow rate of atom loss from these systems
. For instance, the total predicted loss rate for a trapped cloud
of attractive $^{7}$Li atoms of roughly the same number as that in
\cite{Salomon} is around $50/$s.\cite{Pozzi} (This is analogous
to an optical pulse traveling along a fiber with an attenuation of
less than $10^{-6}$dB/km!)

Drawbacks in dealing with atomic clouds include the effect of finite
temperatures. This places a classical uncertainty in the center of
mass momentum and therefore leads to a classical spread in center
of mass position over time, obscuring the quantum diffusion. For a
1D gas of $N$ bosons of mass $m$ and temperature $T$, the equipartition
theorem gives \begin{equation}
\dot{\sigma}_{th}=\sqrt{\frac{k_{B}T}{mN}}\end{equation}
 where $\dot{\sigma}_{th}$ is the linear rate of increase in the
thermal mean position uncertainty.

For example, using numbers from the experiment of Khaykovich \emph{et
al.}\cite{Salomon} and assuming a condensate temperature of $0.5T_{c}\simeq80$nK
we find that the standard deviation in the mean position due to thermal
noise increases linearly at a rate of $1.3\times10^{-4}$m/s. This
is quite fast, when compared to the effect of quantum spreading of
the soliton wave packet, which increases the standard deviation at
the much slower rate of $4.2\times10^{-5}$m/s. As this is an order
of magnitude smaller than the thermal uncertainty, whose growth rate
goes as the square root of the cloud temperature, this implies that
temperatures less than $1$nK are required to directly observe quantum
wave packet spreading. Although this is well below what is achieved
in current $^{7}$Li experiments, Leanhardt \emph{et al.}\cite{Ketterle}
have achieved temperatures of less than $0.5$nK in condensates of
spin-polarized $^{23}$Na, indicating that it may not be impossible
to achieve such low temperatures in matter-wave soliton experiments.
We note that adiabatic changes in the interaction strength could be
used to compress the soliton, thereby increasing the quantum position
uncertainty relative to the standard quantum limit.

\subsection{Degenerate Fermi gasses}

Ultra-cold degenerate Fermi gases have recently been obtained in several
laboratories\cite{formation,Helium3}. The prospects for fermionic
solitons have not yet been established clearly, although tunable Feshbach
resonances with both attractive and repulsive interactions are known
to exist. The species observed experimentally include $^{40}K$ ,
$^{6}Li$ and $^{3}He^{*}.$ Atom correlations\cite{Jin} have already
been experimentally observed in optical measurements with $^{40}K$.
One of the most interesting cases is metastable fermionic Helium,
in which atoms can be counted directly using multi-channel plate detectors\cite{Helium3}.
This affords an interesting possible avenue to testing the prediction
that center-of-mass fluctuations are reduced for fermions relative
to bosons.

\subsection{Double-slit interference of soliton `particles'}

\begin{figure}
\vspace{0.5cm}
 \includegraphics[width=7cm]{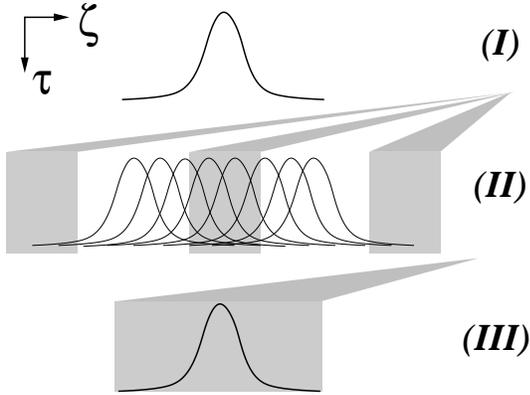}

\caption{\label{fig:doubleslit}Possible soliton double slit interference
experiment. The soliton wave packet (I) broadens over time due to
the linear increase in the quantum uncertainty in the mean position.
Lasers then eliminate solitons outside of two `slits' (II). Any remaining
soliton is left to propagate until it undergoes absorption imaging
(III).}
\end{figure}

Provided one could overcome this difficulty, it would be extremely
interesting to conduct double slit interference experiments with such
large quantum objects. One possible implementation of this idea --
essentially an atomic implementation of the fiber experiment proposal
in \cite{FiniHagelstein} -- is shown in schematic as Fig.~\ref{fig:doubleslit}.
Here a matter-wave soliton is allowed to propagate without disturbance
until its quantum wave packet is large compared with the soliton width.
A tightly focused laser pulse is then used to remove the central and
outer elements of the distribution, effectively creating a pair of
apertures through which the remaining components of the distribution
pass. At some later time, the distribution is imaged and the mean
position of the soliton measured. Another approach - provided the
separation was large compared to the soliton size - would be to use
temporal switching of a localized reflective potential, in order to
only affect one part of the quantum soliton superposition. Averaging
over many shots should then produce an interference pattern with any
thermal effects being manifest as a loss in fringe visibility. 

One of the primary challenges one would face in conducting this experiment
would be addressing the fact that its duration needs to be less than
the average particle lifetime in order to maintain coherence, while
still long enough to allow sufficient quantum wave packet spreading.
A possible solution might be to use a blue-detuned laser to provide
a small repulsive potential near the center of the wave packet to
enhance the broadening and thereby reduce the length of time required
to produce an interference pattern. This technique might also allow
one to forgo the use of destructive laser pulses to create the apertures
by instead relying on this repulsive potential to separate the center
of mass wave packet into two components. Another technique which appears
possible, is to employ the recently proposed soliton quantum beam-splitter\cite{LeeBrand}
to initially separate two quantum soliton components based on their
velocities, prior to subsequent recombination.

A related challenge would be resolving the resulting fringes, as their
width is inversely proportional to particle number and limited by
the trap lifetime through the maximum length of the experiment. (This
can be shown by treating the soliton as a \mbox{de Broglie} wave
of mass $mN$; $m$ and $N$ being the single-particle mass and the
total number of constituent atoms respectively.)

Despite these issues, such an interference pattern would provide smoking
gun evidence that quantum mechanical superpositions of massive composite
objects of mesoscopic scale had been achieved.

\section{Summary}

In summary, quantum wave-packet spreading is a remarkable macroscopic
quantum effect. It provides a fundamental limit - independent of amplifier
noise - both to high-speed communications in dispersive waveguides,
and to applications of atom lasers. Since it provides a means by which
quantum mechanics can couple to the gravitational field, it may also
provide a route to new tests of quantum mechanics. Of more practical
interest is the fact that the standard quantum limit for center-of-mass
measurement variance is highly sensitive to particle statistics. It
is decreased by a factor of $N$ when the particle statistics are
fermionic. This appears testable in atom-counting experiments with
meta-stable $^{3}He$ .

\begin{acknowledgments}
The authors wish to thank both Y.~Yamamoto and A.~Sykes for helpful
discussions. This work was funded by the Australian Research Council. 
\end{acknowledgments}

\end{document}